# LA$^4$SR: illuminating the dark proteome with generative AI

David R. Nelson[1], Ashish Kumar Jaiswal[1], Noha Ismail[1,2], Alexandra Mystikou[1,3], Kourosh Salehi-Ashtiani[1]

1. Laboratory of Algal, Synthetic, and Systems Biology, Division of Science and Math, New York University Abu Dhabi (NYUAD), Abu Dhabi, UAE
2. Department of Biology, New York University, New York, NY, USA
3. Biotechnology Research Center, Technology Innovation Institute (TII), PO Box: 9639, Masdar City, Abu Dhabi, UAE
   †Correspondence should be addressed to D.R.N. (drn2@nyu.edu) or K.S-A. (ksa3@nyu.edu)

**AI language models (LMs) show promise for biological sequence analysis. We re-engineered open-source LMs (e.g., GPT-2, BLOOM, DistilRoBERTa, ELECTRA, and Mamba, ranging from 70 million (m) to 12 billion (B) parameters) for microbial sequence classification. The models achieved F1 scores up to 95 and operated 16,580x faster and at 2.9x the recall of BLASTP. They effectively classified the algal "dark proteome", (e.g., uncharacterized proteins comprising ~65% of total proteins), validated on new data including a new, complete Hi-C/Pacbio *Chlamydomonas* genome. Larger (> 1B) LA$^4$SR models reached high accuracy (F1 > 86) when trained on less than 2% of available data, rapidly achieving strong generalization capacity. High accuracy was achieved when training data had intact or scrambled terminal information, demonstrating robust generalization to incomplete sequences. Finally, we provide custom AI explainability software tools for attributing amino acid patterns to AI generative processes and interpret their outputs in evolutionary and biophysical contexts.**

The application of deep learning to amino acid sequences has the potential to distill fundamental protein features into semantically rich representations encompassing structural, evolutionary, and biophysical properties[1]. Leveraging recent massive expansions in protein sequence databases, large-scale unsupervised learning of protein language models can capture multiscale biological information, enabling state-of-the-art performance in various downstream tasks[2]. In this study, we present the LA$^4$SR (/ˈleɪ.zər/, language modeling with artificial intelligence for algal amino acid sequence representation) framework, based in Pytorch[3], designed to implement the latest open-source language models (LMs) and large language models (LMs with > one billion (1B) parameters, i.e., LLMs) to process microbial genomic data and extract otherwise intractable information.

Microalgal genomes are excellent subjects for this approach[4-6]. The unique niche ecologies of the different algal lineages underscore the diversity of the underlying genetic contents in these microbes, and several phyla display complex evolutionary histories involving lineage-merging endosymbiotic events[7]. Algae genomes are thus notorious for their chimerism[7-9], with many sequences appearing more diverse within a single genome than in other eukaryotes. Further complicating classification is the prevalence of horizontal gene transfer (HGT) in algal genomes[8,10-15]. This evolutionary history blurs the lines between algal and bacterial genetic material, often preventing accurate sequence classification. Traditional bioinformatics tools like BLAST[16] and Kraken[17], which rely on sequence homology and kmer frequencies, often fall short when analyzing novel or divergent sequences. These programs frequently misclassify substantial portions of genuine microalgal genomes as unknown or bacterial. The caveat of incomplete assignment of alignment-based database hits at lower sensitivities also severely hampers the accurate estimation of contamination for genomic sequencing projects. Deep neural networks (DNNs) offer a paradigm shift in sequence data analysis[18], capable of learning complex, hierarchical patterns that enable more nuanced and context-aware classification.

Our novel DNN-based language models show substantial improvements in accuracy, recall, and computational efficiency over traditional bioinformatics methods, such as BLAST, facilitating the scalable, probabilistic phylogenetic mapping of previously unclassifiable sequences. By implicitly incorporating knowledge of algal genome chimerism and horizontal gene transfer events in the training data, we developed a robust classification system for microbial genomic data. The ability to differentiate between algal and bacterial sequences with high accuracy demonstrates the potential of transfer learning in bioinformatics, bridging the gap between general language understanding and specific biological sequence analysis[19]. The new models presented here effectively classify genes comprising the "dark proteome"[20,21], which includes translated gene sets with no hits otherwise returned in alignment-based approaches. In our datasets, the "dark proteome" consisted of 100% [all proteins coded for by a single genome] minus 34.7% [the average percentage of proteins per genome returning BLAST hits from the National Center for Biotechnology Information (NCBI) non-redundant (nr) protein database (https://www.ncbi.nlm.nih.gov/refseq/about/nonredundantproteins/; see Fig. S1] = 65.3% of microalgal proteins with no hits in the nr database. We show that this uncharacterized majority fraction of proteins can be nearly completely classified using the generative AI approaches outlined in LA$^4$SR.

## Results

**Dataset Generation, Model Architecture, and Training Strategies.** We constructed microbial genomics sequence datasets from $n$ = 161 microalgal genomes and known contaminant sequences extracted from nr. Two main methodologies were used for downstream training and evaluation: full-length protein sequences (TI-inclusive) and scrambled start/stop sites (TI-free), allowing us to investigate the role of terminal information (TI) in prediction models (Figs. 1 and S1). These two divergent approaches embodied "core" and "integrated" approaches to elucidate position-independent position-inclusive amino acid patterns inherent to each group. Genomic information was distilled to translated ORFeomes (Data S1). Algal and contaminant sequences were combined for training and evaluation at a 1:1 ratio, with 58,650,525 and 17,880,279 sequences in the TI-free and TI-inclusive training datasets. Overall, ten microalgal phyla were represented in the training data: Chlorophyta, Haptophyta, Cercozoa, Ochrophyta, Dinophyta, Euglenophyta, Heterokonta, Streptophyta, and Chromerida. Combined with the contaminant sequences (i.e., bacterial, archaeal, and fungal sequences), the training data comprised ~77 million distinct sequences.

Our training approach used extracted configurations of foundation models for pre-training (architecture mimicry) and post-training (i.e., fine-tuning) using parameter-efficient fine-tuning (PEFT)[22]. For pre-training, we implemented language modeling approaches using the Hugging Face Transformers[23] library (see Data S2). Our implementation leveraged the HuggingFace Transformers library's AutoModel[24] functionality to streamline model initialization and configuration. For example, LA$^4$SR GPT-NeoX[25]- and Pythia[26]-based models used the GPTNeoXConfig framework, with key hyperparameters tuned for optimal performance. The AutoModel classes provide an automated mechanism for instantiating pre-trained models, handling the complexity of model-specific architectures while maintaining a consistent interface across different model types. Pre-training parameters, including dropout rates and attention priming, were maintained through the configuration object. For example, the LA$^4$SR Pythia 70m architecture was configured with a model dimension of 512, with six layers and eight attention heads, effectively mimicking the carefully researched architecture from this open-source model. All other architectures for our pre-trained models were identical to the base model with uninitialized weights. We post-trained existing models with Low-Rank Adaptation (LORA)[27]. Models with more than 300 million parameters (300m) were post-trained with Quantized LORA (QLORA)[28]. Post-training always started from the pre-trained model weights. The models used in analyses included the S6 Mamba[29,30] and transformer-based Mistral 7B[31], GPT-2 variants (nanoGPT[32] and GPT-NeoX[25]), ByT5[33], DistilRoberta[34-36], MiniLM[37], BLOOM-560m[38], and Pythia (7B and 70m)[26] architectures and pre-trained models. The base architectures and the LA$^4$SR models are available at https://huggingface.co/.

**Model Performance and Technical Efficiency.** We evaluated our models using custom datasets, comparing sequences with and without terminal information (TI-inclusive and TI-free, respectively; Data S1) for algal sequence classification (Figs. 1 and S1). Performance benchmarks, shown in Figure 1, demonstrate the models' capabilities when processing both types of sequences. The model assessment involved rigorous testing on held-out data while tracking computational resource utilization (Figs. 1, S2, and S3). Notably, even in the early stages of training, many post-trained models demonstrated a remarkable baseline understanding of "algal" and "bacterial" concepts in natural language contexts, likely due to their pre-training on scientific literature. This was particularly evident with the Mamba[29,30] 370m model, which was pre-trained solely on The Pile (EleutherAI, https://www.eleuther.ai)—an 800GB diverse text dataset. Remarkably, after just 50 steps of LA$^4$SR post-training, the model could generate responses to bacterial queries by drawing upon scientific literature within The Pile, including substantial content from PubMed (https://pubmed.ncbi.nlm.nih.gov) and GitHub (https://github.com) repositories. This early emergence of domain knowledge suggests efficient transfer learning from the pre-training phase.

Accuracy was validated with new, real-world sequencing data (Fig. S3). The LA$^4$SR framework demonstrated high technical versatility and efficiency across various architectures, ranging in size from 70m to 12B (Figs. 1 and S2). Most transformer-based models (nanoGPT, GPT-NeoX, Mistral, and Pythia) and the S6-based Mamba[29,30] exhibited robust performance with 20,000 – 50,000 training steps using batch sizes of 16 - 96 (i.e., after seeing 320,000 – 4.8m sequences; Fig. 1). Batch sizes up to 512 were tested with success, but training runs with batch sizes larger than 96 usually showed worse accuracy. This result corroborates evidence illustrated by Shirish Keskar et al. (2016)[39] and Shallue et al. (2018)[40] that larger batch sizes can lead to poorer generalization and convergence to sharp minima in the loss landscape, suggesting an optimal batch size window exists below which training may be inefficient and above which accuracy tends to degrade.

Our training experiments showed that batch sizes of 64-96 were ideal for pre-training models smaller than 300m and post-training larger models on an NVIDIA[41] (Santa Clara, CA, USA) A100 GPU, achieving high precision with the lowest runtimes. Notably, larger models (>300m) achieved high accuracy with minimal training data, sometimes using less than 2% of the dataset. A LA$^4$SR Mistral 7B model post-trained (QLORA)[28] with the TI-free dataset achieved F1 scores of ~88 on algal/bacterial classification after only 2,000 training steps (batch size = 96), while pre-training from scratch required roughly 20,000 steps to generate models with F1 scores >80 (Fig. 1). Our pre-training yielded a variety of models that could distinguish between algal and bacterial proteins. The best models had the highest average F1 scores for algal and bacterial input. The TI-free approach had higher inference speeds compared to full-length sequence processing.

In addition to several transformer architectures, LA$^4$SR models were developed with non-transformer S6 LMs. Several recent reports have touted the strong performance and scalability of the non-quadratic-scaling S6 model Mamba[29] for various language tasks. We investigated whether Mamba models could be post-trained for our applications and how scaling could be surveyed and optimized. The resulting Mamba adaptors, comprised of 130m, 370m, 780m, 1.4B, and 2.8B parameter pre-trained models, performed well on classification tasks, achieving F1 scores of up to 88 when evaluating known microalgal and bacterial sequences not used in training. The low/mid-size 370m model showed optimal performance, with large models achieving comparable accuracy but much longer inference times. The mid-size Mamba S6 models were notable for training and inference robustness and demonstrated the strong performance of S6 models as transformer alternatives for AI-based microbial genomics applications. The 370m (Huggingface.co model, "ChlorophyllChampion/Mamba-370m-88F1-45000") and 2.8B (Huggingface.co model "ChlorophyllChampion/LA4SR-Mamba-2.8b-QLORA-ft") LA$^4$SR Mamba models are available for download at huggingface.co.

Compared to BLASTP, we found that LA$^4$SR models can perform at equivalent or better accuracy with much better recall rates and speed. We compared the LA$^4$SR against Diamond BLASTP in terms of accuracy and recall (Fig. 1) as well as technical benchmarks (Fig. S2). We found that LA$^4$SR models perform better than Diamond BLASTP[42], operating an average of 82.90x faster ($P = 4.7 \times 10^{-96}$ on $n = 166$ genomes) and with ~3x the recall rate (~100% in LA$^4$SR compared to ~35% in ultra-sensitive Diamond BLASTP (Table S1)). Thus, the transformer and S6-based modeling facilitated higher recall than alignment-based methods. Although recently undergoing several large-scale sequencing projects, microalgae are still an understudied group with poor representation in sequence databases. The lack of adequate representation hinders database alignment attempts, especially for new sequences. We show the strong capacity of LA$^4$SR models to generalize to "never-seen-before" sequences, demonstrating nearly perfect recall on holdout datasets (Fig. 1a), whole genomes (Fig 1b), and new real-world sequencing data from axenic and xenic lab cultures sequenced with short reads or long reads and chromatin capture methods (Fig. S3, assemblies are in Data S4 and also hosted at NCBI (SAMN44618602)).

The best-performing TI-free model, based on EleutherAI's Pythia[26] 70m architecture (LA$^4$SR -EP70m), achieved balanced and high-performance metrics (precision = 0.9031, recall = 0.8983, and F1 score = 0.9006). Similar results were observed for bacterial sequences, with a precision of 0.8981, a recall of 0.9029, and an F1 score of 0.9003. The model's high accuracy without relying on terminal information (TI) highlights its effectiveness in classifying protein sequences. The strong performance of this small model also provides insights into the fundamental biological differences between these organisms. The compact size of this high-performing model suggests that the primary genetic distinctions between algae and bacteria can be captured by a limited number of parameters. This indicates a potentially low-rank problem where the essential differences can be represented in a lower-dimensional space. Such a low-rank structure indicates interpretable features that could be translated into human-understandable concepts.

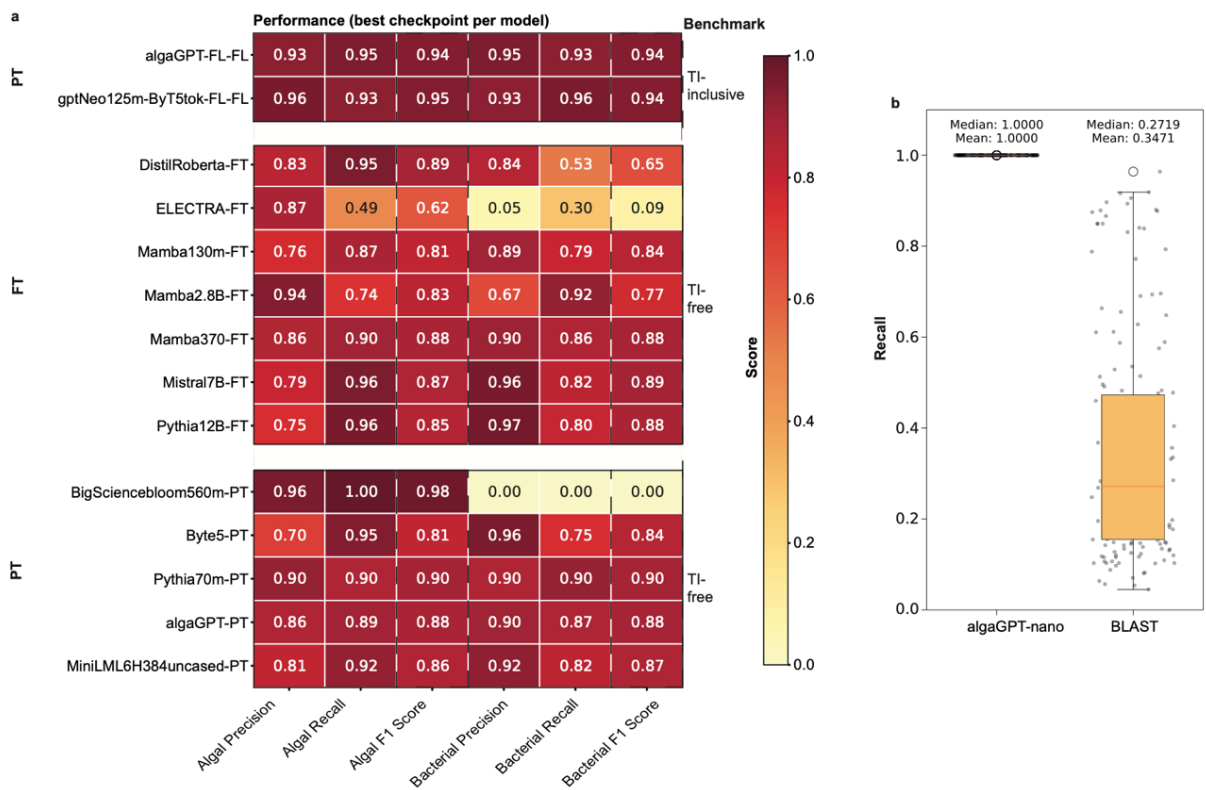

**Fig. 1 | Performance comparison of language model architectures for amino acid sequence classification. a,** Pre-training (PT) from scratch with LM architectures produces high-accuracy, general-purpose amino acid sequence models. The best-performing checkpoint was used to create the final pre-trained model. Fig. S1 shows further details of the experimental design, and Fig. S2 shows performance metrics from different checkpoints for selected models. Finetuning (FT; i.e., post-training) existing pre-trained models also achieved high accuracy, and often in fewer training steps. The heatmap shows the accuracy of models pre-trained from scratch using various LM architectures for algal amino acid sequence classification. **b,** Comparison of recall rates for ultra-sensitive Diamond BLASTP[42] and algaGPT on translated coding sequences from $n$ = 166 microalgal genomes[43].

Our comprehensive performance analysis demonstrates a revolutionary advancement in sequence alignment capabilities using LA$^4$SR. When evaluating traditional alignment software, represented by BLASTP[16] and Diamond BLASTP[42], we found that for a standard comparative task—aligning a query sequence against a database of 1 million sequences (averaging 400 amino acids)—an A100 GPU could theoretically outperform a high-end CPU core by a factor of approximately 1,560. However, LA$^4$SR models, which encode comprehensive training data knowledge within their weights, achieve dramatically superior performance beyond this theoretical GPU acceleration.

Our rigorous benchmarking reveals that LA$^4$SR models not only match BLASTP's accuracy but substantially exceed it, operating at an unprecedented speed increase of 16,580x while achieving 2.9 times higher recall rates (100% in LA$^4$SR versus 34.7 ± 0.25% in Diamond BLASTP's ultra-sensitive mode). Most significantly, these models successfully address one of bioinformatic's most challenging problems: the classification of the "dark proteome"[20,21], —those translated gene sets that yield no hits through conventional alignment-based approaches. In our microalgal datasets, this dark proteome constitutes a substantial 65.3% of proteins (calculated as the difference between the complete proteome [100%] and proteins returning BLAST hits from the NCBI non-redundant protein database [34.7%; see Fig. S1]). LA$^4$SR achieves near-complete classification of this previously uncharacterizable majority fraction.

Implementing our TI-free approach further enhanced performance, enabling token generation an order of magnitude faster (Fig. S2). These breakthrough results represent more than incremental improvements; they constitute a paradigm shift in high-throughput bioinformatic analyses. By combining GPU acceleration, sophisticated machine learning architectures, and optimized data processing pipelines, LA$^4$SR demonstrates that deep learning approaches can transcend the limitations of traditional sequence alignment methods. This advancement not only accelerates existing analytical capabilities but presents new possibilities for understanding previously uncharacterizable protein sequences, potentially revolutionizing our ability to explore and understand protein diversity across life's domains.

**Model Interpretability and Feature Analysis.** What sequence features and molecular patterns drive our high-performing models' decision-making process? To answer this fundamental question, we ran multifaceted model interpretability[44] analyses, probing deep into the neural networks' learning mechanisms. Our investigation leveraged several leading interpretation techniques, implementing custom explainer programs (Data S3) and illuminating how models process and evaluate sequence information. Through the strategic application of advanced interpretability frameworks—including Tuned Lens[45], Captum[46], DeepLift47[47], and SHAP-based approaches—we systematically decoded how specific amino acid residues, their patterns, and positional relationships influence model decisions. This analytical approach led to the development of sophisticated model-agnostic gradient explainer tools (see Data S3 and Fig. S4), enabling us to extract and visualize crucial gradient information from multiple perspectives, such as per-residue, per-position, and per-motif attribution scores.

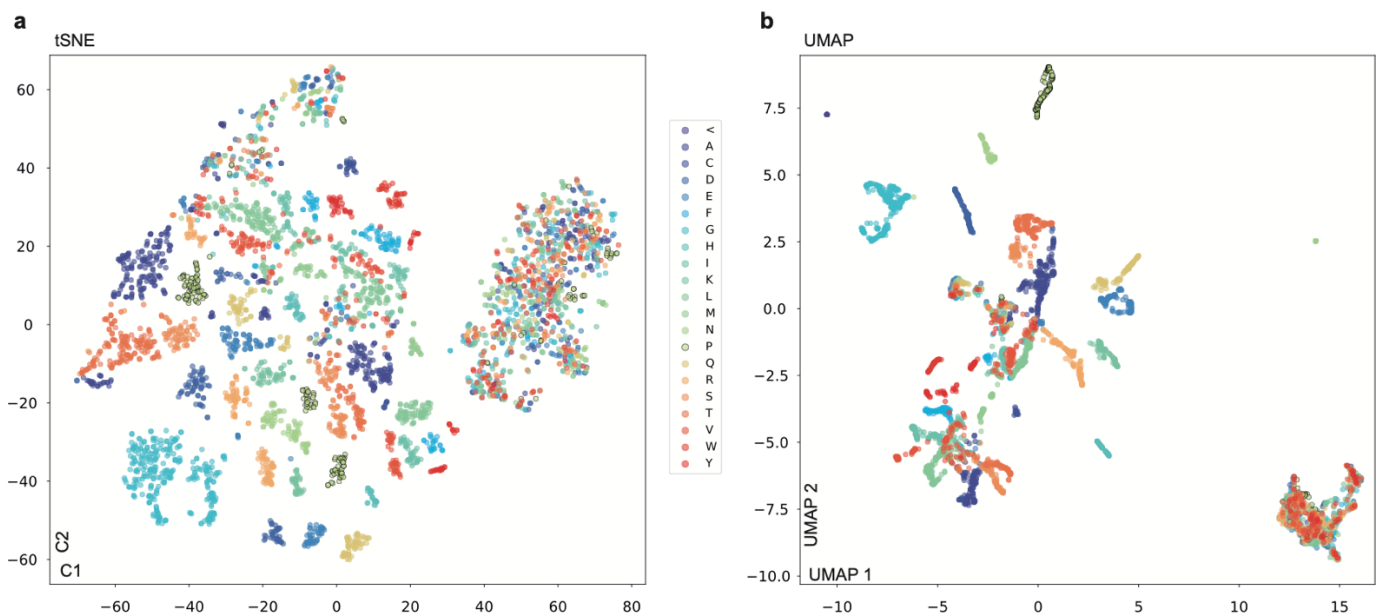

**Fig. 2 | Dimensionality reduction techniques applied to integrated gradient attribution scores from a Pythia[26] 70m-based LA⁴SR model processing randomly selected bacterial sequences. a,** T-stochastic neighbor embedding (t-SNE)[48] visualization of sequence-level attribution patterns. **b**, Uniform manifold approximation and projection (UMAP)[49] clustering. Points are colored by amino acid type. While t-SNE emphasizes local structure and cluster separation, UMAP better preserves global relationships. The attributions were computed at the token level (where each token represents a single amino acid). This visualization reveals how sequences with similar attribution patterns cluster, suggesting the model has learned consistent patterns of sequence importance across the bacterial dataset.

We saw that glutamine and glycine dominated token usage frequency (Fig. 3). The disproportionate emphasis on glycine and glutamine in distinguishing between eukaryotic algae and bacteria highlights the importance of amino acid usage patterns in reflecting evolutionary adaptations to different ecological niches, particularly in relation to nitrogen metabolism. This may be due to the roles of these amino acids in signaling the assimilation and accumulation of nitrogen, represented by glutamine, and its restricted cellular availability, represented by glycine[50-52]. The availability of nitrogen limits growth in both groups, although through substantially different metabolic pathways. Thus, these metabolic differences may have been implicitly learned in this LA⁴SR instance as key features discriminating the algal and bacterial classes. Valine showed higher usage in model decisions when algal sequences were sampled, and tyrosine was more frequently used for bacterial queries (Fig. 3). These two amino acids showed an inverse relationship in the way these two amino acids are used when discriminating algal and bacterial sequences.

The newly developed explainer programs presented here (HELIX, DeepLift LA⁴SR, and Deep Motif Miner Pro [DMMP]) implement distinct approaches to transformer hidden state interpretation. HELIX performs layer-wise principal component analysis (PCA) transformations on hidden states (e.g., Pythia 70m's dimension $d = 768$) to analyze embedding space topology. Each hidden state matrix $H \in \mathbb{R}^{(n \times d)}$ (where n = sequence length) undergoes orthogonal transformation to identify principal axes of variation in the embedding space, enabling tracking of residue relationships across network depths. DeepLift LA⁴SR quantifies feature importance through attribution relative to a zero-reference state. The method processes sequences in windows ($w = 32$, stride = 16) computing attribution scores $A = \sum(DL(x, x_0))$, where DL represents the DeepLift operation comparing input x against baseline $x_0$. The implementation wraps the embedding layer $E: \Sigma^* \to \mathbb{R}^d$ for direct gradient access, with attribution scores summed across embedding dimensions: score(i) = $\sum_k A(i,k)$ for position i and embedding dimension k. The DMMP software calculates position-wise influence scores through relative hidden state distances. For each amino acid at position i, it computes $I(a,i) = \|\mu_a - \mu_{(\sim a)}\|_2$, where $\mu_a$ is the mean hidden state for residue a and $\mu_{(\sim a)}$ is the mean state for all other residues. The method identifies motifs M = {(s,p,v)} where s is a subsequence starting at position p with influence value v exceeding the 95th percentile threshold $\theta = P_{95}(\{I(a,i)\})$. DMMP aggregates layer information through concatenation H' = [$H_1$;...;$H_l$] followed by dimensionality reduction using PCA, t-SNE, or UMAP matrices. The methods' core differences lie in their attribution calculations: HELIX uses purely geometric transformations, LA$^4$SR employs backpropagation-based attribution against reference states, and DMMP utilizes distance metrics in hidden state space combined with statistical thresholding. LA$^4$SR processes sequences through sliding windows maintaining ω=16 position overlap, while HELIX and DMMP operate on full sequences with batch-wise memory management.

A custom Captum[46] analysis pipeline (Data S3) evaluated feature contributions in LA$^4$SR's amino acid sequence classification tasks. This Captum-based layer-integrated gradients analysis includes various model-agnostic visualization tools for exploring amino acid patterns as features in language models. We computed attributions for each input token, quantifying their impact on the model's predictions. For each input sequence, normalized attribution scores and model prediction probabilities were calculated, providing insights into which parts of the input most strongly influenced the model's decision-making process. The end-of-sequence token and the two to five preceding tokens, representing terminal information (TI), had higher influence scores than sequences from other region trends (Fig. 4b,c). This trend was observed in all transformer models tested except for the bidirectional DistilRoBERTa[34-36] post-trained model. Still, tokens at the ends of sequences had higher influence scores in the DistilRoBERTa LA$^4$SR model.

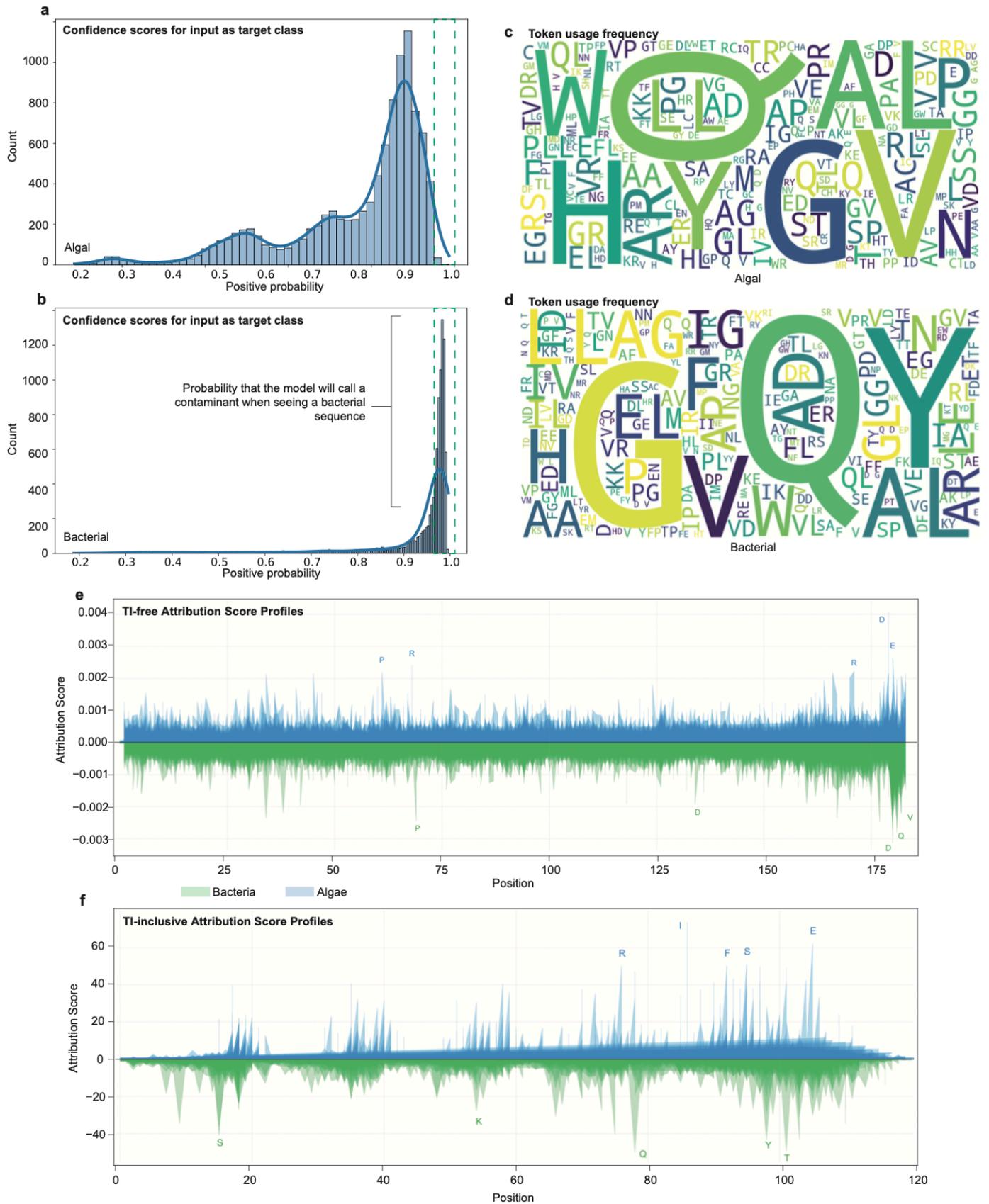

**Fig. 3 | Model attribution analysis for sequence classification. a,** Confidence score distribution from Captum[46] attribution analysis of a GPT-NeoX-125m-based pre-trained LA$^4$SR model, evaluating the classification of algal sequences. The histogram shows positive probability scores (0-1) across 20,000 TI-free representative algal and bacterial sequences, with higher values indicating

stronger classification confidence. The distribution demonstrates a multi-modal pattern with a primary peak at high confidence (>0.9) and secondary peaks at lower confidence levels, suggesting varying sequence-specific certainty. **b,** Corresponding probability distribution for bacterial test sequences (n=10,000), showing a pronounced peak near 1.0, indicating a high-confidence classification of bacterial contaminants. The sharper, more concentrated distribution compared to algal sequences suggests more distinctive bacterial sequence features. **c,** Word cloud visualization representing token usage frequencies for algal inputs. **d,** Word cloud showing token usage frequences for bacterial inputs. **e,** DeepLift[47] attribution score profiles for TI-free sequences, analyzed using a sliding window approach (window=32 residues, stride=16 residues) across 10,000 sequences per class. Shaded fits are at ~10% of the most conservative estimates. The balanced positive (blue) and negative (green) attribution signals demonstrate position-independent feature extraction, characteristic of a TI-free architecture. **f,** Attribution profiles for TI-inclusive model showing pronounced periodic patterns and cumulative attribution effects correlating with sequence length. The increased magnitude and periodicity of attributions reveal position-dependent feature learning, distinguishing the TI-inclusive model's sequence processing from its TI-free counterpart. Letters above significant peaks indicate key discriminative residues for classification decisions.

**Influential Motif Identification.** We developed the DMMP software (see Data S3) to identify influential motifs in the model and quantify their impacts on model decisions. This method analyzes hidden states from the model and calculates position- and residue-specific influence scores across all layers to identify motifs that strongly influence the model's decisions. We used a multi-step process to identify influential amino acids within protein sequences, leveraging the hidden state representations from the GPTNeoX[53]-based LA$^4$SR model. Our approach involved extracting hidden states for each input sequence, computing influence scores by calculating the Euclidean distance between the mean representation of each amino acid and the mean of all others, normalizing these scores, and then using peak detection to identify local maxima. For each layer's hidden states, we computed a position-wise influence score using the L2 norm of the difference between each position's representation and the mean representation of that layer: influence = $\|h\_i - \mu\|\_2$, where $h\_i$ is the hidden state at position i and μ is the mean hidden state for the layer. This operation is vectorized across all positions: position_influence = np.array([np.linalg.norm(state - np.mean(state, axis=0), axis=1) for a state in hidden_states]). The resulting scores were min-max normalized to a [0, 1] range, then across all layers to obtain a single influence score per position. To identify peaks in this influence profile, we applied the find_peaks function from scipy.signal with a height threshold set at the 95th percentile of all scores and a minimum peak distance of positions. Around each identified peak, we extracted a window of amino acids to form potential motifs. Patterns were extracted that exert a disproportionate influence on the model's decision-making process, revealing class-defining signatures. These peaks are considered the centers of influential motifs, extracted as subsequences of three to six residues centered on each peak. The motifs are then ranked by their average influence score, providing a prioritized list of potentially functionally or structurally important amino acid subsequences within the protein.

In approximately 15% of algal sequences classified as bacterial-like, we identified specific motifs (e.g., "GXGKT" and "DXXG") associated with ATP/GTP-binding sites (Fig. 4, Table S4). These motifs showed higher-than-average influence scores, suggesting their strong impact on classification decisions and potentially indicating horizontal gene transfer events. To visualize and interpret these results, we used several techniques, including amino acid representation plots using PCA, influence score plots across layers, and additional dimensionality reduction techniques such as t-SNE and UMAP. These visualizations provide different perspectives on how the model clusters amino acids and its understanding of them evolves through the network.

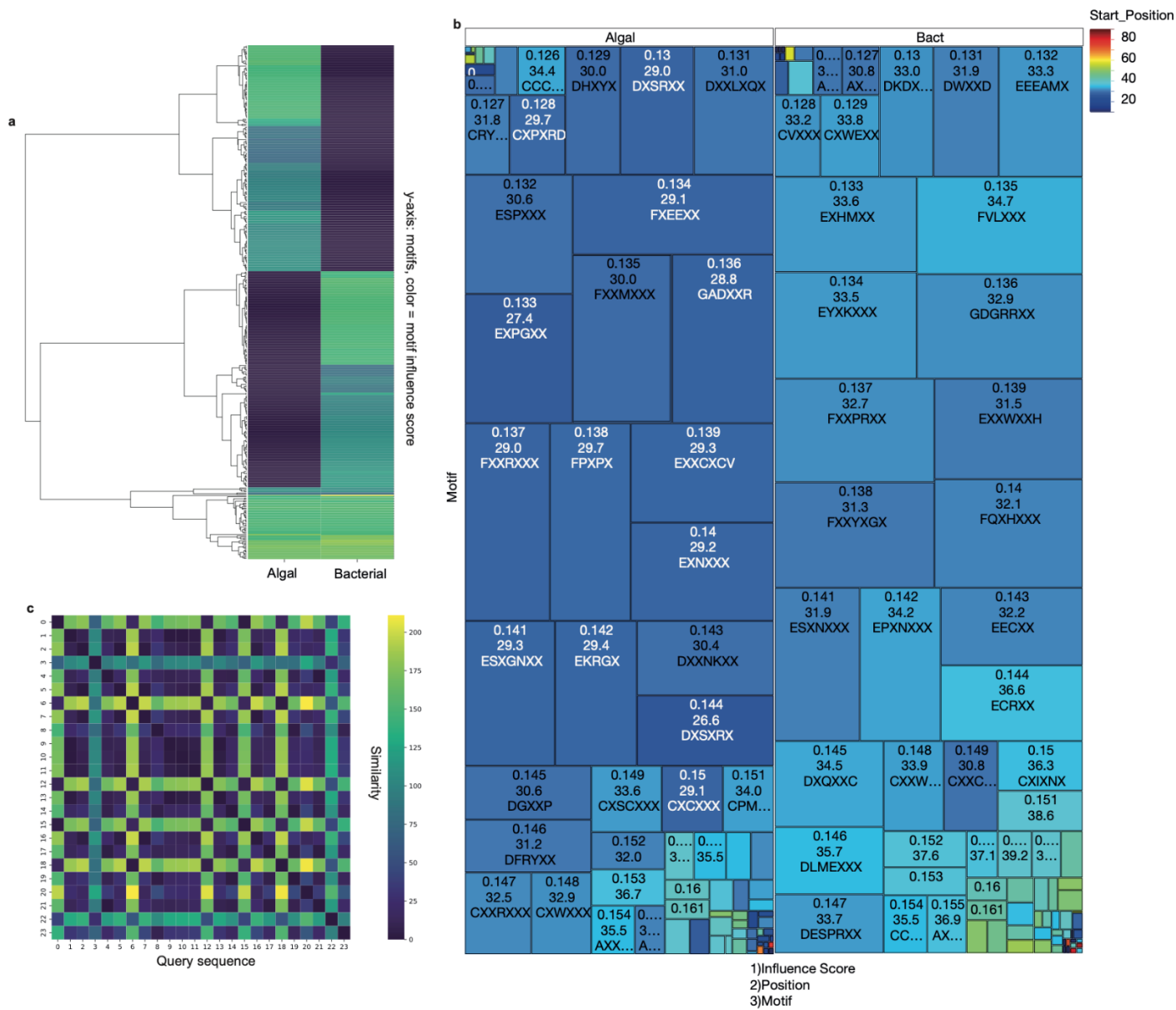

**Fig. 4 | Examining sequence influence in LA⁴SR with motif attribution for next-token prediction using MotifMiner Pro.** A GPT-NeoX[25] architecture-based pre-trained LA⁴SR was used to process amino acid sequences and extract hidden state representations across multiple transformer layers. Then, flexible motif identification was applied. The suite includes functionality to visualize amino acid representations in 2D space, analyze the influence of individual amino acids across model layers, and generate comprehensive heatmaps of flexible motif influences across multiple sequences. Altogether, 10,000 sequences for each of the algal and bacterial groups were analyzed; aggregative statistics are shown. **a,** A heatmap displaying the influence scores of various motifs in the query sequence. Motifs are categorized as either algal or bacterial based on their ground truth origin and not model determination. **b,** Influential motif compositions and positions. Algal motifs are, in general, slightly earlier in the average query sequence, which suggests an accelerated understanding of algal sequence patterns compared to bacterial sequence patterns. The text in white shows motifs starting earlier than 30 residues. **c,** Sequence similarity plot; sequences are the first 23 lines from the shuffled training set where the ground truth is known. The crosshatch pattern shows similarity among motif distributions in the training algal and bacterial sequences used by the model to differentiate the two groups.

To complement the Captum approach and expand the capacity for explainer tools to non-causal LMs, such as the DistilRoBERTa[34-36]-based LA⁴SR model, we developed a SHAP (SHapley Additive exPlanations)[54]-like gradient explainer to decipher sequence attribution (Data S3). This software implements a custom gradient explainer to compute attribution values directly from PyTorch[3] output. The program quantifies the contribution of each amino acid to the model's classification decision and elaborates on features influencing the model (Fig. S4). Our CustomShapExplainer class calculates these values by backpropagating from the loss of predicting specific tag tokens to the input embeddings, overcoming the limitations of traditional methods when applied to large language models.

Our SHAP-like value analysis of the DistilRoBERTa[34-36]-based LA$^4$SR model revealed class-specific patterns in its interpretation of sequences. Algal sequences displayed diverse attribution scores in SHAP-like attribution value heatmaps (visualized as varying shades of blue; Fig. S4). In contrast, bacterial sequences showed more uniform patterns (predominantly colored in red). This consistency across bacterial samples indicates that the model has identified robust features characteristic of bacterial proteins. At the same time, diversity in algal attributions suggests a more complex, nuanced understanding of algal sequence features.

**Layer-wise, per-residue and per-position attributions.** To gain deeper insights into the internal representations of LA$^4$SR models, we implemented layer-wise analyses with Captum[46]'s IntegratedGradient library and our own custom PyTorch[3] method to probe intermediate representations in transformer models. Our custom implementation (Data S3, HELIX (Hidden Embedding Layer Information eXplorer)) extracts intermediate representations from each layer of LA$^4$SR models and formats them for downstream visualization. HELIX extracts hidden state vectors from each transformer layer and visualizes projections in lower-dimensional space. The visualization tool plots amino acid token embeddings with consistent markers and colors across layers, allowing direct observation of how sequence representations evolve through the model's depth. The resulting plots reveal clustering patterns and spatial relationships between amino acid embeddings at each layer (e.g., Fig. 5).

HELIX (Fig 5; Data S3) revealed patterns in the model's processing of amino acid sequences. We observed an increasing trend in representation complexity for most amino acids up to the fifth layer, followed by a convergence in the final layer, outlining a refinement of features as information propagates through the network (Fig. 5). By the third layer, distinct clusters of amino acids began to form, possibly corresponding to shared physicochemical properties or functional roles in protein structures. These cluster formations revealed that the model has learned biologically relevant groupings. The final layer showed the most diverse spread of amino acid representations, suggesting highly specialized features for each amino acid that contribute to the model's classification decisions.

The distribution of influence scores in unidirectional transformer models, exemplified by the LA$^4$SR Pythia model, showed a subtle inverted Gaussian-like pattern, with a slightly concave shape visible in the upper envelope of the data points (Fig. 6b,c). This curvature is characterized by slightly higher values at the extreme ends of the x-axis (i.e., termini) compared to the middle. The overall shape shows a dense concentration of points at lower y-values, with a gradual upward spread. From left to right along the x-axis, an increase in the vertical spread of points was observed, indicating greater variability in measured values at higher start positions. Instead of perfectly even spreads, our data reveal these influence score distribution trends as inherent artifacts of these causal models (Fig. S4). The custom Captum implementation connected model decisions to human-understandable concepts, including group-specific fingerprints or "spectra" from influence signatures in amino acid sequences (Figs. 3, 4; Table S3).

The distinctive trends we observed that were unique to cysteine and methionine are peculiar considering the evolutionary significance these residues represent. These two sulfur-containing amino acids exhibited unique representation patterns across layers (Fig. 5), potentially reflecting their special roles in protein folding and redox cellular processes. Methionine consistently appeared at the center of amino acid clusters in layer-wise projections in PCA plots, aligning with its known function as the initiator of protein synthesis. Notably, cysteine, which also contains sulfur, exhibited atypical trends in these PCA plots, occupying peripheral projections. This anomaly may be linked to cysteine's crucial role in redoxome enzymes, which are fundamental to a species' antioxidant capacity[55-57]. Given that antioxidant capabilities significantly influence evolutionary trajectories[58-60], these distinctive cysteine influence patterns could serve as markers of lineage divergence. Traditional phylogenomic methods have struggled to elucidate the complex interplay between environmental pressures and lineage relationships. Meanwhile, our results suggest that deep neural networks can capture and unravel these intricate, long-term interactions. This machine learning-driven approach opens new avenues for understanding the molecular basis of evolution, potentially bridging the gap between sequence-level variations and macro-evolutionary trends.

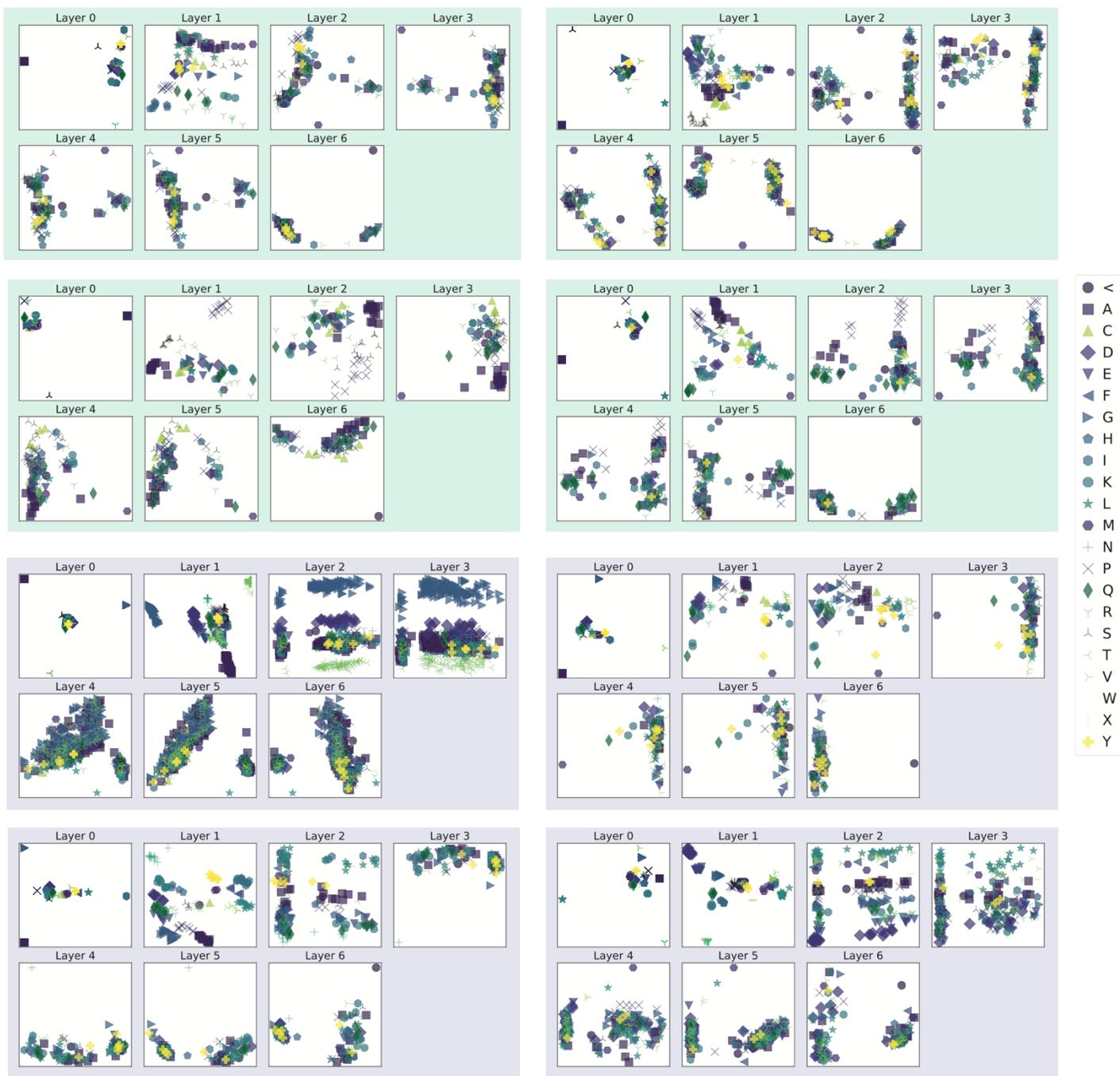

**Fig. 5 | Layer-wise evolution of feature representations in a TI-free Pythia-based LA$^4$SR model.** The progressions of amino acid feature representations through the layers of a deep learning model trained on algal (green) and bacterial (blue) genomic data as detected in HELIX (Data S3) are shown. The scatter plots depict the distribution of amino acid features in 2D space after dimensionality reduction with PCA for each layer of the neural network. Each point represents an amino acid. Each scatter subplot shows performance at an individual layer from analysis on a single algal sequence. This visualization demonstrates the neural network's transformation of genomic data into discriminative features based on their per-residue attribution scores. Fig. S7 shows per-position features. The increasing separation in deeper layers reflects the model's growing ability to distinguish between genomic signatures in both groups. The consistent presence of all classes across layers indicates maintained information throughout the network's depth. Expanding axis ranges in deeper layers suggest feature space expansion for improved separation. The figure visualizes feature representations across seven neural network layers (Layer 0 to Layer 6). Colors and shapes denote different residues as indicated in the legend. Layer 0 (Input Layer): a dispersed distribution with some clustering but significant class overlap. Layers 1-3: progressive feature separation, indicating learning of algal sequence type differentiation. Layers 4-5: further refinement of feature clusters and multi-dimensional separation. Layer 6 (Output Layer): the most pronounced separation suggesting refined features.

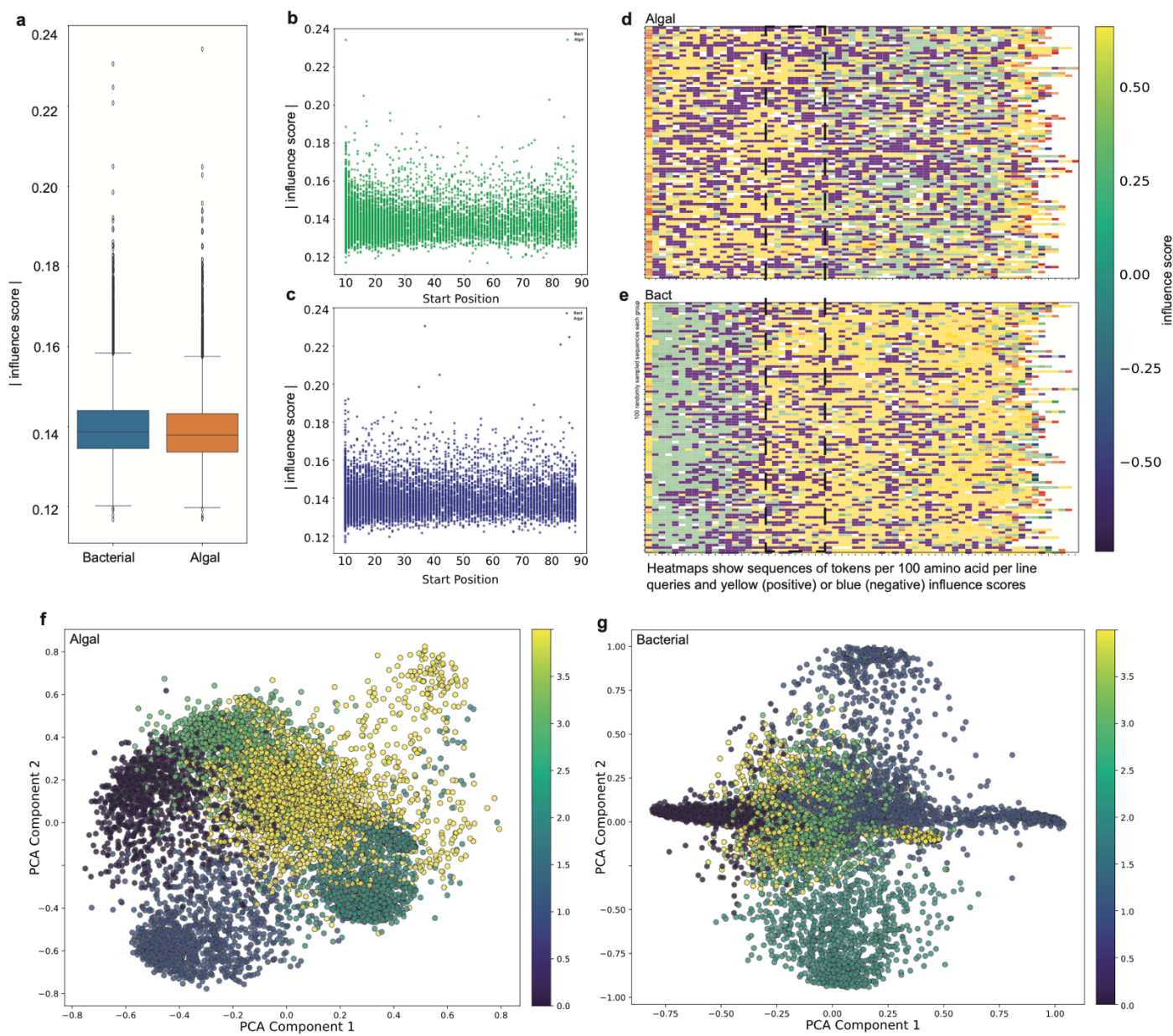

**Fig. 6 | Captum[46] analysis of algal and bacterial per-residue attribution scores. a,** Distributions of absolute influence scores in algal and bacterial sequences in a Pythia 70m-based LA$^4$SR model. **b-c,** Distributions of absolute influence scores per position in algal and bacterial sequences, showing a relatively even distribution of influence scores nearly independent of position. **d-e,** Heatmaps showing influence scores from sequences of tokens per 100 amino acid queries for algal (top) and bacterial (bottom) samples. The x-axis represents the position in the sequence, and the y-axis represents individual samples. The color scale ranges from -0.50 (blue: negative influence scores) to 0.50 (yellow: positive influence scores). The contrasting heatmaps give insight into the LA$^4$SR prediction process. The sequences around the 30 amino acid mark appear highly influential in the Pythia-based LA$^4$SR model. This finding was further supported by our motif-based analyses (Fig. 4), where the most influential three-to-six amino residue motifs were usually found 29-34 residues into the sampled sequence. **f-g,** Principal Component Analysis (PCA) scatter plots for Captum attribution scores from algal (left) and bacterial (right) input sequences ($n = 10,000$ sequences each class; see Table S3). The color (z) scale represents the absolute influence, where z=|attribution score|. Each point represents a sample or sequence.

**Validation with real-world data.** To validate our approach and address real-world challenges, we applied LA$^4$SR models to new data from new clean and contaminated isogenic algae cultures. This application was crucial for differentiating between natural algal sequences, which can sometimes appear bacteria-like, and true contaminant species—a distinction vital for accurately assessing the genetic makeup of algal samples and identifying potential contamination in sequencing data. We cultured and sequenced ten separate isogenic colonies of *Chlamydomonas reinhardtii* CC-1883. Of these, nine were sequenced with Illumina 150 bp paired-end short reads and one with Pacific Biosciences (PacBio, Menlo Park, CA, USA) HiFi reads and DoveTail (Sydney, Australia) Hi-C to generate a complete, axenic reference assembly (Fig. S3). Two axenic *Bigelowiella natans* cultures sequenced independently from geographically distant labs, and published in different studies[7,43] yielded very similar genomes with relatively high counts of bacterial-like sequences as determined using LA$^4$SR as well as BLASTP[16]. Out of 9,749 and 9,571 total queries with BLASTP hits, 1005 and 1035 sequences had the highest number of hits in bacteria (Table S2). These results (Fig. S3 and Table S1) bring the maximum expected *bona fide* algal sequences with bacterial sequences per genome up to 10.56 ± 0.36%.

Overall, BLAST detection of contaminants suffered from a lack of sensitivity or indecisiveness, with no hits found for 65.3 ± 0.25% of input sequences from all genomes tested ($n = 166$) with ultra-sensitive Diamond BLASTP runs (E = 0.00001, see Table S1). For reference, the contamination removal pipeline BLEACH[61] was named because of its occasionally harsh and indiscriminate flagging and removal of false positive contaminants for the tradeoff of increased true-positive contaminant removal. In contrast to BLASTP, algaGPT-nano made predictions for > 99% of sequences in each input genome (Table S1). Similar near-complete predictions could be achieved with most LA$^4$SR models on most genomes (~ 99%). The rate of predicted contamination to algal phylogenetic assignment was comparable in BLASTP and LA$^4$SR, although LA$^4$SR predictions generally called *bona fide* algal sequences as such, whereas BLAST would call them a contaminant (Fig. S3). In conclusion, the LA$^4$SR AI models demonstrated performance equivalent to or surpassing that of conventional CPU-based methods in contaminant detection. They achieved approximately three times the recall rate while operating at speeds orders of magnitude faster. Preceding performance gains, this work shows the successful application of a broad range of open-source AI software to niche, high-level custom biological questions.

**Discussion**

Our study's application of next-generation language models (LMs) to biological sequence analysis represents a significant advancement in computational biology. This approach highlights the potential of transfer learning in bioinformatics, bridging the gap between general language understanding and specific biological sequence analysis. The investigation into terminal information (TI) in LA$^4$SR models' decision-making algorithms revealed intriguing parallels with human language processing. The observed reliance on TI, even when scrambled, mirrors cognitive science findings about the importance of word beginnings and endings in human language processing. This reveals potential similarities between artificial and biological neural networks in sequence interpretation, opening new avenues for interdisciplinary research at the intersection of computational biology and cognitive science. However, the heavy reliance on TI also exposes a potential limitation in current transformer-based models. The bias toward terminal sequences might lead to overlooking important internal patterns, which is particularly problematic for incomplete or poorly annotated genomes. Our development of TI-free models addresses this issue, demonstrating comparable performance without relying on terminal information. This approach not only improves the robustness of LA$^4$SR models but also challenges the field to reconsider the design of attention mechanisms in biological sequence analysis.

The transfer learning paradigm presented in this work capitalizes on the vast amount of knowledge embedded in the model design and the pre-trained weights, significantly reducing the required task-specific training data and enhancing performance on downstream tasks such as phylogenetic classification. The models' pre-existing understanding of complex language structures translates remarkably well to the "language" of protein sequences, enabling the capture of subtle patterns and relationships that might be overlooked by traditional methods, especially when dealing with highly divergent or novel sequences. A major focus of the study was investigating the effect of terminal information (TI) in the decision-making algorithms of the models we developed. Deep neural networks, as well as human brain neural networks, weigh the ends of words more than the intermediate letters when deciphering meaning from text[62-64]. This phenomenon is reminiscent of the interactive activation model proposed by McClelland and Rumelhart, which accounts for context effects in letter perception[65]. The importance of terminal information has been observed across languages, as demonstrated in studies of Chinese word recognition[66]. Research on non-contiguous letter combinations further supports the significance of letter position in word processing[62,64]. Drawing parallels from speech recognition, where recurrent neural networks have shown remarkable performance in processing sequential information, we considered how similar principles might apply to protein sequences.

The concept of long-term dependencies in sequential data, as explored in the development of long short-term memory (LSTM) networks[67], provided a theoretical foundation for our investigation into the importance of terminal information in protein sequences. Additionally, spatial coding models of visual word identification offered insights into how position-specific information is processed in LA$^4$SR neural network models. We hypothesized that the models would also place disproportional importance on sequences near the end of proteins. This would be problematic for genomes with incomplete annotation because these end regions would not be properly resolved. The TI-free models were designed to circumvent any detrimental effects from TI by scrambling TI so that the model would not receive this information during training. The results provide key insights into how these open-source LMs work with amino acid sequence queries and how different patterns and not sequence lengths characterize these microbes' genomes. Developing models that can effectively balance the interpretation of both terminal and internal sequence features is crucial for many use cases.

The success of our TI-free approach suggests that robust, functionally relevant patterns can persist even when peripheral data are altered or removed, which could be particularly valuable in analyzing fragmentary or poorly assembled genomic data. This finding could lead to new strategies for dealing with incomplete or noisy biological data, a common challenge in genomics and proteomics. The multi-faceted interpretability framework presented here, combining gradient-based, perturbation-based, and layer-wise analyses, enhances our ability to interpret and validate predictions. As AI models become increasingly complex, such interpretability methods will be crucial for building trust and extracting meaningful biological insights. The increasing complexity in deep learning models often results in "black box" systems that, while powerful, lack transparency and interpretability[68]. This opacity can hinder trust, adoption, and the extraction of meaningful biological insights. The LA$^4$SR models leverage state-of-the-art language modeling techniques to analyze algal amino acid sequences, focusing on interpretability and causal inference. This approach not only predicts and classifies but also elucidates the underlying mechanisms of sequence pattern learning—a crucial feature in bioinformatics, where understanding predictions is as vital as making them.

Although we incorporated more than 77 million distinct sequences in the training data, the biological sequence data available for several representative groups remains sparse. Despite efforts to include diverse algal and bacterial sequences, an underrepresentation of less sampled algal lineages, including Chromeridia, Dinoflagellates, Rhodophytes, Ochrophytes, and, in general, non-Chlorophyte or Bacillariophyta genomes in public databases, limits the scope of the work. These challenges present an opportunity for innovation in database and training data generation and the development of new model architectures and training methodologies, potentially leading to more sophisticated and biologically relevant machine learning models. In conclusion, our work with LA$^4$SR not only advances bioinformatics through improved sequence analysis but also opens new research directions at the intersection of machine learning, cognitive science, and biology. The insights gained from comparing TI and TI-free approaches could inform future model designs, potentially leading to more robust and versatile tools for analyzing the complexities of biological systems.

This work contributes to improving the accuracy of algal genome assembly and annotation, enhancing our understanding of microbial evolution and biodiversity, and supporting biotechnological applications that rely on pure algal genetic material. As computational biology continues to evolve, the integration of language models with existing methods and experimental approaches, as exemplified in this study, will be critical in advancing the understanding of the molecular languages of life on Earth, particularly in complex and diverse microbial communities such as those found in algal ecosystems. Combining powerful deep learning models with robust explainability techniques paves the way for more transparent, reliable, and biologically meaningful microbial genomics analysis.

## Methods

**Open-source model architectures.** The LA⁴SR framework integrated several open-source software packages and models. We used low-rank adaptation (LORA)[27] and quantized low-rank adaptation (QLORA) for parameter-efficient post-training and Mamba as an alternative to transformer[69]-based architectures. The Hugging Face (https://huggingface.co) transformers library facilitated implementation and post-training of the open-source models, while Bitsandbytes enabled quantization and memory-efficient training. For model analysis and interpretability, we developed custom gradient extraction programs based on SHAP[54,70] (SHapley Additive exPlanations, https://github.com/shap/shap), Captum, and DeepLift. We post-trained large language models including Mistralai/Mistral-7B-v0.1 (https://huggingface.co/mistralai/Mistral-7B-v0.1), as well as various Mamba pre-trained models from the Hugging Face model hub. We used the S6 Mamba[29,30] and transformer-based Mistral[31], GPT variants (nanoGPT[32], GPT-NeoX[53]), ByT5[33], DistilRoBERTa[34-36], MiniLM[37], BLOOM-560m[38], and Pythia (7B and 70m)[26] for pre-training and post-training.

Each model had an idiosyncratic methodology, but uniformity in coding was achieved using AutoModel[24]. This system automatically handled model-specific configurations through the from the _pretrained() method, which loads pre-trained weights from either local storage or HuggingFace's model hub. The initialization process involved three main components accessed through their respective Auto classes (AutoConfig, AutoModel, and AutoTokenizer), maintaining consistent configurations, model weights, and tokenizers while preventing component incompatibilities. The AutoModel approach facilitated the architecture-agnostic implementation of LA⁴SR, where model-specific details are abstracted away through the AutoConfig class. This class automatically selected appropriate configurations based on the model type, handling architecture-specific parameters such as model dimensions, the number of layers, and attention heads according to each pre-trained model's specifications.

The DistilRoBERTa[34-36] architecture was the only bidirectional model as well as the only model using sentence transformers (https://sbert.net/). PyTorch was used as the primary deep learning framework used for implementing custom datasets, data loaders, and training pipelines. Additionally, we developed custom Python scripts for data preprocessing, model training, and analysis, including a novel motif-based explainer pipeline for model interpretability. Training was performed on an HPC cluster using NVIDIA[41] (Santa Clara, CA, USA) A100, V100, and H100 GPUs.

Implementing parameter-efficient post-training techniques, such as LORA and QLORA, proved crucial for post-training larger LLMs. The low rank approach allowed us to efficiently fine-tune large models. The relatively low level of input data needed to make the LLMs perform well (compared to sub-1B models, i.e., LMs) suggests that our primary classification task (ie., algal v. bacterial) is a low-rank problem with a small number of extremely complex and important features that can be rapidly learned by LLMs as shown in our post-training of Mistral 7B (Figs. 1 and S2). These influential features are the proteomic evolutionary adaptations characterizing organisms from each group.

**Training Data Preparation.** Our previous large-scale microalgal genomics project, ALG-ALL-CODE[43], generated a large quantity of new algal genome sequences, and these sequences comprised the basis for identifying, through next-token prediction, whether a query sequence is algal. A typical application of microalgal genomics includes the survey of populations, and it is of utmost importance to screen for possible contaminants. Of these, bacteria are the most common but fungal, and to a lesser extent, archaeal species may have been included but undetected during sequencing. The contaminant sequences used for training and testing were obtained from the National Center for Biotechnology Information (NCBI; ncbi.gov) non-redundant (nr) database using accession lists for species in the different phylogenies.

We translated ORFeomes from >100 microalgal genomes from diverse phylogenies and known contaminant (e.g., bacterial, fungal, and archaeal) sequences extracted from the non-redundant (nr) NCBI (https://www.ncbi.nlm.nih.gov) protein database. Algal and contaminant (either mixed with fungi and archaea or purely bacterial) sequences were combined to form the training/evaluation split training datasets at a 1:1 ratio, with 58,650,525 (6,375,974,800 characters) and 17,880,279 (4,224,680,663 characters) sequences in the TI-free and TI-inclusive training datasets (Data S1, https://zenodo.org/records/13920001). Translated ORFeomes (tORFeomes) from approximately 100 microalgae species were first screened for contaminants using traditional BLAST-based detection, and tORFeomes passing a 10% filter were maintained. This threshold was chosen based on results we obtained from two independently sequenced[7,43], *bona fide* Chloroarachniophyte genomes. The greatest quantity of bacteria-like sequences that could be predicted from known clean algal genomes from the BLAST[16]/BLEACH[43] approach was from the Chloroarachniophyte species (*Bigelowiella natans*)[7,43], at about 10% (Table S2). Passing genomes were divided into training, evaluation, and test sets based on the training regime (terminal information [TI] or TI-free) used. Our study examined the influence of sequence length on the decision algorithm in both pre-trained and post-trained models and explored how token masking was applied across different model configurations.

The training data-preparation workflow consisted of large-scale protein/amino acid sequence processing and LLM training, with several distinct processing stages. Sequence preprocessing was performed using a custom pipeline (see Data S2, https://zenodo.org/records/13920001) that handled segmentation, labeling, and format standardization. The core training infrastructure included distributed training scripts for multi-GPU setups alongside specialized training implementations for different model architectures. Post-training capabilities were implemented using fastLoRA-train.py and configurations specified in Data S2. We trained and inferenced models with and without TI (Fig. 1). One class of datasets used full-length sequences (TI-inclusive), while the other scrambled start and stop site information (TI-free, Fig. S1). These two approaches were drastically different, each presenting different caveats and advantages. For full-length sequences, a '<' was added to the end of each sequence to indicate the switch to sample for a label instead of a protein sequence. However, in testing, labels could usually be generated with sequence prompts alone without the need for additional characters. The TI-free training data were generated by first removing all headers and newlines (\n), resulting in one long line of sequence. This unisequence was printed at 100 characters per line, then shuffled—effectively removing all TI. This format allows for the same length to be used as the training sequence, which benefits training and inference procedures, including eliminating the possibility that the model learning length is a factor in its decision algorithm. It further challenges models to predict features not associated with TI, including position-independent amino acid patterns inherent to each class (i.e., core signatures).

One caveat in the TI-free approach was that chimeric, non-natural sequences are produced. However, any chimera would still be in the same group (e.g., algal or bacterial) and was expected to retain group affinity information at sub-full length levels. Shuffled 100-aa sequence files were tagged with a label to report group affinity, concatenated into a master training dataset, and then shuffled again. Under these two classes, training and testing datasets comprised splits of algal/bacterial and algal/bacterial/fungal/archaeae, each with a ratio of algae-to-contaminant of 1:1. We implemented a custom PyTorch Dataset class to preprocess data, defining len and getitem methods for efficient indexing and iteration. The PyTorch DataLoader was then configured with optimized parameters: batch sizes ranging from 32 to 128 sequences, shuffling enabled for improved model generalization, multiple worker processes (4-8) for asynchronous loading, and pinned memory for faster GPU transfer.

Prior to model input, sequences underwent character- or subword-level tokenization using predefined vocabularies, conversion to PyTorch tensors, and uniform length padding within batches. The 100-character format yielded several advantages: it allowed efficient memory allocation, aligned well with our model's architectural design (particularly the attention mechanisms in transformer-based components), and facilitated faster data loading by minimizing additional preprocessing during batch preparation. This carefully optimized data-Prior to model input, sequences underwent character- or subword-level tokenization using predefined vocabularies, conversion to PyTorch tensors, and uniform length padding within batches. The 100-character format yielded several advantages: it allowed efficient memory allocation, aligned well with our model's architectural design (particularly the attention mechanisms in transformer-based components), and facilitated faster data loading by minimizing additional preprocessing during batch preparation. This carefully optimized data handling pipeline significantly enhanced both training efficiency and model performance on the TI-free dataset, demonstrating the critical role of data preprocessing in large-scale sequence analysis tasks. We post-trained the models using learning rates ranging from $1 \times 10^{-4}$ to $1 \times 10^{-5}$, batch sizes of 8 - 96, and context windows (max_seq_length) of 128 - 4096. These parameters were found to yield satisfactory results according to downstream evaluations.

**Tokenization and Model Versatility.** A key challenge in training and evaluating different LM architectures is differentiating performance variations from different sources that may not be inherent to the target models' architectures, such as the tokenization scheme. While LA[4]SR nanoGPT[32] and GPT-NeoX[53]-125m were initially trained on full-length sequences with their default tokenizers, which use a hybrid approach of byte-level tokens and larger subword tokens created through byte-pair encoding (BPE) and fixed vocabularies, pre-training from scratch with other open-source architectures was incompatible with some of the methods in these tokenizers. Furthermore, using a different tokenizer for each model would have precluded accurate downstream comparison for some tasks. For most pre-training operations, we used an adaptable, pure byte-level tokenizer to transcend limitations owing to model-specific functions used in training and inference pipelines. In the context of amino acid sequences, which frequently include rare or modified amino acids, the ByT5 tokenizer could handle any Unicode string without out-of-vocabulary issues. This byte-level approach ensured that all amino acids, regardless of their frequency or complexity, were represented with uniform granularity (see Fig. 2).

This consistency in sequence representation likely enhanced the model's ability to learn and generalize patterns across diverse protein sequences.

We found that the ByT5[71] tokenizer, which uses a vocabulary of just 384 tokens representing individual bytes, across various model architectures, including GPT-NeoX[53] (125m; "EleutherAI/gpt-neo-125m"), had several advantages for protein sequence modeling. The tokenizer's byte-level approach provided consistent representations across different model types without needing model-specific preprocessing. Using the ByT5[71] tokenizer during pre-training across various models eliminated confounding factors introduced by model-specific tokenizers. Its integration with GPT-NeoX[53], ELECTRA[72], and other model infrastructures we tested that are typically associated with their own tokenizers, underscores the ByT5[71] tokenizer's cross-architecture compatibility. Our results show that pure byte-level tokenization can be effectively incorporated to pre-train various open-source architectures (Fig. 1). This approach offers a truly language-agnostic and universally applicable solution for sequence representation in protein-modeling tasks.

**Generative Modeling for Algal and Contaminant Sequence Discrimination in Genomic Data.** Our primary objective was to leverage generative modeling techniques to differentiate between algal and contaminant genomic sequences, where the predominant contaminants come from various bacterial lineages. This task is crucial in algal genomics, where the accurate identification of source material is essential for downstream analyses and applications. Models under 300m parameters were trained from scratch by extracting or mimicking the underlying architecture configuration from the Huggingface (https://huggingface.co/) model, and post-trained models used the weights from the pre-trained models to start. According to our BLAST results that identified substantial portions of *bona fide* algal sequences as bacteria-like (up to ~10% in some species; see Table S1), we did not expect the precision of LA$^4$SR models to reach higher than 90% when inferencing on algal sequences (Table S2). However, in several instances, this metric was surpassed (Fig. 1). Reasons for the superior performance of some LA$^4$SR models compared to the BLAST-based baseline estimates include biases in the input sequence and biases in the determination of the 90% algal precision threshold.

For testing, we used either batches of 1,000, 10,000, or ~300 million sequences per set, representing quick-check, standard, and exhaustive testing datasets not used in training corresponding to either TI-inclusive and TI-free algal, bacterial, fungal, and archaeal sequences not used in training. We reason that sensitivity to learning outlier algal sequence examples by the model determines a domain of accuracy specific to clades with members showing more chimeric genomes, such as dinoflagellates, Chromeridia, Pelagophyta, and Chloroarchniophyta. While the BLAST-based contaminant detection analysis registers ~ 10% contamination in the *bona fide B. natans* proteome, models were trained with all its sequences labeled as algal sequences. Thus, discriminating models based on higher diversity sequences can help inform more holistic decisions regarding lineage.

Well-performing models were expected to achieve high bacterial precision because, in general, bacterial genomes do not contain algal-like sequences. *Bona fide* bacterial sequences called "algal" by the model represent false negatives that may have possibly been learned due to their similarity to bacterial-like algal sequences. These results represent the heaviest caveat of LA$^4$SR; ideally, the stringency of *bona fide* bacterial sequence calls should be paramount. Still, some of the LA$^4$SR models approximate the target stringency at the cost of slightly lower algal precision (Fig. 1). We used mixed-precision training (FP16) with gradient checkpointing to optimize memory usage. We included the AdamW optimizer with a weight decay of 0.01 and an initial learning rate of $1 \times 10^{-4}$ coupled with a cosine learning rate scheduler incorporating 2,000 warmup steps.

The implementation includes robust error handling for invalid model configurations, missing or corrupted model weights, incompatible tokenizer configurations, and resource constraints during model loading. This ensures system stability and provides meaningful error messages for debugging and maintenance. All models present a unified interface for forward passes, state management (training/evaluation modes), and weight loading and saving despite their architectural differences. When necessary, the configuration can be modified for specific downstream tasks while maintaining architectural compatibility, providing flexibility for various applications while preserving the integrity of the underlying model architecture.

Training was run for 0.02 – 3 epochs with a per-device batch size of 16-96 and gradient accumulation steps of 8-32 to simulate larger batch sizes while managing memory constraints. To monitor training progression, we integrated Weights & Biases (wandb.ai) logging, tracking key metrics including training and validation losses. The amount of training to achieve optimal results (F1 > 90) was generally around this level for 100-200m parameter models. In contrast, the smaller Pythia 70m needed longer training times and more input data and the larger models (> 300m) needed less training on average. The high F1 scores for both algal and bacterial sequences achieved in many LA$^4$SR models indicate that they have learned to capture the distinctive features of each group, making it a valuable tool for microbial community analysis and protein function prediction.

Balancing algal and bacterial precision was not necessarily a function of the input training dataset size or training time. For example, a GPT-NeoX-125m model trained with a ByT5 tokenizer exhibited a "sweet spot" of model performance at 30,000 training steps (batch size = 64), while the 25,000- and 35,000-step checkpoints showed lower performance. At the 25,000-step checkpoint, the model showed good performance, with an algal F1 score of 0.7646 and a bacterial F1 score of 0.6741. The algal precision was notably high at 0.8871, indicating a low false positive rate for algal sequence identification, but here, the recall was low at 0.6720. The model's performance improved significantly by the 30,000-step checkpoint. The algal F1 score increased to 0.8712, with a remarkable improvement in recall (0.9602) while maintaining a good precision (0.7975). The bacterial F1 score also increased to 0.8903, with high precision (0.9666) and good recall (0.8253). This checkpoint demonstrated a well-balanced performance for both algal and bacterial sequence classification.

Our GPT-NeoX-125m training instance exemplified the trend of LMs often reaching diminishing returns and even degraded performance with extended training regimes. Although recall was maintained, precision wavered after 30,000 steps. The 35,000-step checkpoint showed a slight decrease in performance compared to the 30,000-step mark. The algal F1 score was 0.8688, with a precision of 0.8008 and a recall of 0.9496. The bacterial F1 score was 0.8855, with a precision of 0.9566 and a recall of 0.8244. Similar diminishing returns or performance drops were commonly observed in overtrained LA$^4$SR models. The 30,000-step checkpoint model showed the overall best performance on the TI-free benchmark testing set, suggesting that this might be an optimal point for model selection. Thus, exported models were always taken from checkpoints with the highest performance on the testing benchmark datasets (Fig. 1). Models are hosted at Huggingface.co as described in Table S1.

**Technical evaluations of LA$^4$SR and traditional architectures.** To assess the computational requirements of LA$^4$SR models, we calculated several key metrics. These included the number of parameters (Params), multiply-accumulate operations (MACs), floating-point operations (FLOPs), and floating-point operations per second (FLOPS). We distinguished between forward propagation FLOPs (fwd FLOPs) and backward propagation FLOPs (bwd FLOPs), assuming that the default model backpropagation requires twice the computation of forward propagation. We calculated these metrics for both the forward pass alone (fwd) and combined forward and backward passes (fwd + bwd). The total training parameters were computed, along with fwd MACs, fwd FLOPs, fwd +bwd MACs, and fwd+bwd FLOPs. All values were calculated and reported in appropriate units (M for millions, G for billions).

The larger models we tested reached higher accuracy with less training data. For instance, a QLORA-post-trained Mistral 7B model reached an F1 score of ~88 on algal/bacterial classification after only 2,000 training steps, compared to roughly 20,000 steps required for smaller models trained from scratch. The TI-free approach, using fixed-length 100-amino-acid sequences, had higher inference speeds than full-length sequence processing. Memory optimization techniques, including quantization and gradient accumulation, enabled efficient training and inference even on limited hardware. This scalability and flexibility make LA$^4$SR adaptable to various research environments, from individual workstations to high-performance clusters, offering a powerful tool for microbial genomics and protein sequence analysis across different computational settings.

One of the primary motivating factors for our work was instilling the capacity to work with the latest GPU technologies[41] into bioinformatics software. The A100 GPU demonstrated a 200-400× improvement in floating-point operations per second compared to single-core CPU performance in double-precision calculations. For example, the performance gap between a GPU, specifically the NVIDIA[41] (Santa Clara, CA, USA) A100, used here in most calculations, and a typical high-performance CPU core for a BLASTP-like algorithm is monumental and is due to several technical factors. At its core, this difference stems from the fundamental architectural disparities between these two types of processors. The A100 GPU has 6,912 CUDA cores capable of massive parallelism, delivering up to 19.5 TFLOPS in double precision and 156 TFLOPS in single precision. This raw computational power, coupled with a staggering 1.6 TB/s of memory bandwidth, stands in stark contrast to a typical high-performance CPU core, which might offer 50 - 100 GFLOPS and 50 - 100 GB/s of memory bandwidth. This translates to the GPU potentially performing 200 - 400 times more floating-point operations per second than a single CPU core in double precision, and even more in single precision. We tested speed using translated coding sequences from whole algal genomes (n = 166) as queries for ultra-sensitive Diamond BlastP and the LA$^4$SR model algaGPT (Table S1). The algaGPT model was 82.9x faster, on average, than ultra-sensitive Diamond BlastP. Genomes generally finished in under one hour with algaGPT, while Diamond BLAST runs took 1-2 days using 28 cores on an Intel processor. The sensitive mode of Diamond BLAST was previously found to be 2,000x faster[42] than NCBI BLAST; thus, using NCBI BLAST for these genome-level comparisons was not feasible. We thus used the estimate from Buchfink et al.[42] to compare the speed of our models to NCBI BLAST, translating to a 16,580x speedup of algaGPT compared to NCBI BLAST.

Our analyses revealed large variations in both computational requirements and potential runtime performance. Large language models (i.e, > 1 billion (B) parameters) demonstrated the highest computational intensity. For example, the Mistral7B-1000 model could run at 1830 GFLOPS and 912.8 GMACs, coupled with the largest parameter count of 7000m (7B), suggesting it would likely have the longest runtime among the models studied. Conversely, the 19m TI-free LA$_4$SR model with a Pythia 70m architecture had the lowest computational needs (5.09 GFLOPS and 2.54 GMACs) and a compact parameter count of 19.31m, indicating potential for faster execution. The sentence-transformers-all-distilroberta-v1-dualityFT50000 and Byt5-Prophecy100s-30000 models exhibited similar computational profiles (approximately 22 GFLOPS and 11 GMACs), suggesting comparable runtimes, despite slight differences in their parameter counts (82.12m and 86.14m). The 70m TI-free LA$_4$SR model with Pythia 70m architecture occupied a middle ground regarding terms of both computational intensity and potential speed, with 11.63 GFLOPS, 5.81 GMACs, and 70.43m parameters. Notably, the TI-inclusive LA$_4$SR model based on gpt-neo-125m architecture showed moderate computational requirements (32.24 GFLOPS and 16.11 GMACs) but had a larger parameter count of 125.2M, suggesting a balance between model size and runtime efficiency. These metrics provide insights into the trade-offs between model complexity and potential execution speed, crucial factors in selecting models for various applications with different performance requirements.

For a more granular analysis, we broke down the calculations for each module in the model. For each module, we reported its parameters, percentage of total parameters, MACs, percentage of total MACs, FLOPS, and percentage of total FLOPs (Table S2). Some modules may use torch.nn.module or torch.nn.functional to compute logits (e.g., CrossEntropyLoss), which are not counted as submodules in our calculations. This can result in a discrepancy between a parent module's MACs and the sum of its submodules' MACs. Additionally, we acknowledge that the number of floating-point operations is a theoretical estimation, which may result in calculated FLOPS exceeding the maximum system throughput.

**Real-world Sequencing Validation.** The foundational datasets we used for training consisted of real-world sequencing experiments, either at the whole-genome or large-scale, whole-genome level. Here, to validate LA$_4$SR models with new real-world data, we performed both short-read and long-read sequencing on additional axenic and xenic algal cultures. Ten isogenic colonies of *Chlamydomonas reinhardtii* strain CC-1883 (cw15 NIT+ mt-; (https://www.chlamycollection.org/product/cc-1883-cw15-nit-mt/) were isolated and cultured under standard growth conditions. Nine colonies were sequenced using Illumina NovaSeq 6000 with 150 bp paired-end reads. For the tenth colony, we generated a complete reference assembly using a combination of PacBio HiFi long reads and Hi-C chromatin capture data. Cultures were maintained in Tris-Acetate-Phosphate (TAP) medium at 25°C under continuous light (100 μmol photons m$^{-2}$ s$^{-1}$) with constant shaking at 150 rpm. Genomic DNA was extracted using an protocol optimized for algal cells as indicated in Nelson, et al., 2024[61]. For Illumina sequencing, DNA libraries were prepared using the NEBNext Ultra II DNA Library Prep Kit following the manufacturer's instructions. Paired-end sequencing was performed with a target coverage of 50X per sample. For the PacBio HiFi sequencing, high molecular weight DNA was extracted and size-selected for 15-20 kb fragments. Hi-C libraries were prepared using the Dovetail Omni-C Kit following the manufacturer's protocol. This produced a chromosome-level assembly that served as a clean, axenic reference. For validation, we also analyzed previously published genomes from isolates of *Bigelowiella natans,* known for containing bacterial-like sequences due to their evolutionary history[7], from two separate whole-genome sequencing projects. We evaluated potential contamination using three approaches: traditional BLAST-based detection using the BLEACH pipeline, ultra-sensitive Diamond BLASTP (E = 0.00001), and LA'SR models, including algaGPT-nano. Our analyses showed that several of the genomes would not finish after one month's runtime with NCBI BLASTP, effectively precluding further investigation with this tool for our purposes. Ultra-sensitive Diamond BLASTP commonly finished jobs in approximately two days or less, and most LA$_4$SR jobs finished in under one hour's runtime.

The results from axenic cultures established baseline false positive rates for bacterial-like sequence detection. The analysis of the *B. natans* genomes revealed that up to 10.561 ± 0.357% of bona fide algal sequences may present bacterial-like characteristics, providing an important benchmark for model evaluation. This reference point helped calibrate expectations for LA$_4$SR model precision, particularly when analyzing organisms with complex evolutionary histories. BLASTP showed limited sensitivity, failing to classify 65.3 ± 0.25% of input sequences from the tested genomes (n = 166) even in the ultra-sensitive mode. In contrast, LA$_4$SR models achieved near-complete sequence classification (~99%) while maintaining accuracy comparable to or exceeding BLAST-based methods. The real-world validation demonstrated that LA$_4$SR models could effectively distinguish between genuine bacterial contamination and algal sequences with bacterial-like characteristics, a crucial capability for accurate genome analysis. The newly sequenced genomes shown in Fig. S3, including a new Hi-C/PacBio reference assembly for the *C. reinhardtii* strain CC-1883, are available in Data S4 (https://zenodo.org/records/13920001) and at NCBI (SAMN44618602).

**SHAP-like gradient explainer.** SHAP (SHapley Additive exPlanations) represents a unified framework for interpreting predictions by combining game theory with local explanations. At its core, SHAP uses Shapley values from cooperative game theory to assign importance values to each feature for a particular prediction. It has a strong theoretical foundation, providing consistent and locally accurate explanations that satisfy important properties like local accuracy and missingness. SHAP stands out for its model-agnostic nature, meaning it can explain any machine learning model's output, and its ability to provide both local (individual prediction) and global (entire model) interpretations. However, SHAP's computational complexity can be significant, especially for large datasets or complex models, as calculating exact Shapley values requires evaluating all possible feature combinations. We re-engineered the SHAP pipeline to suit our models (SHAP-like Explainer; Data S3, https://zenodo.org/records/13920001). SHAP-like values were implemented in our work using our DistilRoBERTa[34-36]-based LA$_4$SR model as an example.

The "SHAP-like Explainer" (see Data S3) introduces a new class "CustomShapExplainer" and two new functions (explain(self, input_data, target_tags) and shap_explain(model, tokenizer, test_dataset, num_samples=n)) to extract SHAP[54] -like values for input sequences. Representative output from the custom SHAP-like extraction method is shown in Fig. S4. Briefly, the method calculates loss towards a target (e.g., the next token) and calculates SHAP-like values with shap-like_values = (input_embeds.grad * input_embeds).sum(dim=-1).detach().cpu().numpy() using the gradients from the input embeddings to designate the influence towards the "algal" or "bacterial" decision. We demonstrated its application by analyzing how a DistilRoBERTa[34-36] -based LA$_4$SR model processes algal and bacterial queries (Fig. S4).

**Captum workflows.** Captum (Meta AI) offers a comprehensive library of attribution algorithms specifically designed for PyTorch models. It implements various interpretation methods, including integrated gradients, layer conductance, neuron conductance, and several other gradient-based approaches. Captum's main advantage is its deep integration with PyTorch, making it particularly valuable for deep learning practitioners using this framework. The library excels in providing detailed insights into neural network behavior, allowing attribution at different levels—from individual neurons to entire layers. However, Captum's specialization in PyTorch can be a limitation for users working with other frameworks, and its focus on gradient-based methods may not be ideal for all models or analysis needs.

To analyze the model's token attributions, we used the LayerIntegratedGradients and DeepLift methods from the Captum[46] library. For each input text, we first tokenized the input and used the model to generate nine new tokens. We then applied LayerIntegratedGradients to compute attributions for the entire generated sequence. The algorithm used 50 integration steps and zero tensors as baselines for both input IDs and attention masks. This process allowed us to quantify the importance of each input and generated token in influencing the model's final token prediction. The resulting attributions were normalized and visualized using a color-coded sequence, with green representing positive attributions and red representing negative ones. This visualization, along with the original text, generated text, and computed tag probabilities, was compiled into a report for each input sample.

To identify distinct patterns in the attribution scores across protein sequences, we performed a k-means clustering analysis. The optimal number of clusters was determined using the elbow method, which examines the relationship between the number of clusters (k) and the within-cluster sum of squares (inertia). For each value of k from 1 to min(10, n), where n is the number of sequences, we calculated the percentage decrease in inertia compared to k-1. The optimal k was selected when adding additional clusters yielded diminishing returns, defined as the first k where the percentage decrease in inertia fell below 20%. Prior to clustering, the attribution scores were standardized using z-score normalization to ensure equal weighting of features. The resulting clusters were visualized using Principal Component Analysis (PCA). This approach allowed us to identify natural groupings in the attribution patterns while avoiding overfitting through excessive data partitioning.

**DeepLift Explainer analyses.** This analysis used DeepLIFT[47] (Deep Learning Important FeaTures), a method for computing feature importance scores. While DeepLIFT forms the basis for DeepSHAP, it is distinct from Shapley value calculations. DeepLIFT works by comparing neuron activations to reference activations and backpropagating the differences to input features. In the analyses shown in Fig. 3, a sliding window approach was implemented where each window underwent independent tokenization and attribution analysis, with adjacent windows sharing 16-residue overlapping regions (e.g., Window 1: positions 0-31, Window 2: positions 16-47, etc.). This windowing approach enabled fine-grained attribution mapping across the sequence length while maintaining computational efficiency suitable for larger models. The overlapping nature of the windows provides redundant sampling at boundary regions, enhancing the robustness of the attribution signals. Contrasting results from TI-inclusive and TI-free methods showed a distinct position-dependant trend in the TI-inclusive model that was mostly absent

**Transformer layer analyses.** This program registers output from the individual transformer layers. For data manipulation and numerical operations, we used NumPy[73]. Signal processing tasks, particularly peak detection, were handled using SciPy[74] (scipy.signal). We leveraged the Transformers library from Hugging Face (HF: https://huggingface.co/docs/transformers/; Github: https://github.com/huggingface/transformers), including AutoTokenizer and AutoConfig. Model weights were accessed and loaded using the huggingface_hub and safetensors libraries. For dimensionality reduction, we used Scikit-learn's[75] PCA, t-SNE[48], and UMAP[49] from the umap-learn library. Data visualization was accomplished using Matplotlib[76] and svgwrite[77] for scalar vector graphic (SVG) generation. Additional utilities included the gc module for memory management, argparse for parsing command-line arguments, and collections for efficient data structures. The main function for the motif-based analysis program was identify_influential_motifs() (identify_influential_motifs(hidden_states, sequence, window_size=n, percentile=n)). Influence scores were calculated for each position, where position_influence[layer, pos] = np.linalg.norm(pos_state - other_state). To minimize terminal sequence effects, we normalized positional influence scores (position_influence = (position_influence - position_influence.min()) / (position_influence.max() - position_influence.min())). To extract motif-based influence scores, the average influence across layers was calculated (avg_influence = np.mean(position_influence, axis=0). From average influence, peaks (peaks = find_peaks(avg_influence, height=threshold, distance=window_size) above a threshold (threshold = np.percentile(avg_influence, percentile) were returned with the following function: def print_influential_motifs(motifs, avg_influence, percentile, top_n=10): print(f"\nTop top_n Influential Motifs:"); print("Motif | Start Position | Influence Score"); print("-" * 40); for motif, start, score in motifs[:top_n]: print(f"motif:5 | (start:15} | (score:.4f}"). The full scripts are available in Data S3.

**Data availability.** Details on data used for the training and validation of LA$^4$SR models, including sample size and role in the machine learning process, are included in the Supplementary Information. The training and testing datasets as well as the processed new, real-world sequencing data (i.e., Data S1 and S4), are available at Zenodo DOI: 10.5281/zenodo.13920001 (https://zenodo.org/records/13920001).

**Code availability.** All scripts used in this study are available in Data S2 and Data S3, as described in the Supplemental Information, and hosted at Zenodo DOI: 10.5281/zenodo.13920001 (https://zenodo.org/records/13920001).

**Acknowledgements.** We thank members of Salehi-Ashtiani and Jean-Claude Twizere labs for discussions on the project. This research was supported by NYUAD Faculty Research Funds (AD060). This work was carried out on the High Performance Computing resources at New York University Abu Dhabi.

**Author contributions.** D.R.N. wrote the manuscript with input from the other authors and performed formal analyses. K.S.-A. and A.K.J. performed data curation and formal analyses. A.M. grew and prepared the *Chlamydomonas reinhardtii* CC-1883 cw15 NIT+ mt- strain for sequencing; K.S.-A., D.R.N, and A.K.J. developed the conceptual framework of the project and K.S.-A. oversaw the completion of the work.

**Competing interests.** The authors declare no competing interests.